\documentclass[final,10pt,twocolumn,letterpaper,conference]{IEEEtran}

\usepackage{amsmath,amsfonts,amssymb,amsbsy,url,verbatim}
\usepackage{mathtools}
\usepackage{times}
\usepackage[english]{babel}
\usepackage[dvips]{graphicx}
\usepackage{bm}
\usepackage{comment}
\usepackage{stfloats}
\usepackage{cite}
\usepackage{psfrag}
\usepackage{dsfont} 
\usepackage{url}
\usepackage{wrapfig}
\usepackage{algorithm}
\usepackage{algcompatible}
\usepackage{boxedminipage}
\usepackage{multicol}

\newcommand{\bi}{\begin{itemize}}
\newcommand{\ei}{\end{itemize}}
\newcommand{\ben}{\begin{enumerate}}
\newcommand{\een}{\end{enumerate}}
\newcommand{\bc}{\begin{cases}}
\newcommand{\ec}{\end{cases}}
\newcommand{\bd}{\begin{description}}
\newcommand{\ed}{\end{description}}

\newcommand{\be}{\begin{equation}}
\newcommand{\ee}{\end{equation}}
\newcommand{\bea}{\begin{eqnarray}}
\newcommand{\eea}{\end{eqnarray}}

\newcommand{\back}{\!\!\!\!\!}

\graphicspath{%
     {./figs/}%
}

\IEEEoverridecommandlockouts

\begin{document}


\title{Context Information Based Initial Cell Search for Millimeter Wave 5G Cellular Networks}

\author{Waqas Bin Abbas, Michele Zorzi \thanks{The work of Michele Zorzi was partially supported by NYU Wireless.}\\
    \IEEEauthorblockA{Department of Information Engineering, University of Padova, Italy \vspace{0.5mm}}\vspace{-4mm} \\
    \IEEEauthorblockA{E-mail: \texttt{\small \{waqas,zorzi\}@dei.unipd.it} \vspace{-4.5mm}
}}

\maketitle

 \begin{abstract}
 Millimeter wave (mmWave) communication is envisioned as a cornerstone to fulfill the data rate requirements for fifth generation (5G) cellular networks.
In mmWave communication, beamforming is considered as a key technology to combat the high path-loss,
and unlike in conventional microwave communication, beamforming may be necessary even during initial access/cell search.
Among the proposed beamforming schemes for initial cell search, analog beamforming is a power efficient approach but suffers from its inherent search delay during initial access. 
In this work, we argue that analog beamforming can still be a viable choice when context information about mmWave base stations (BS) is available at the mobile station (MS).
We then study how the performance of analog beamforming degrades in case of angular errors in the available context information.
Finally, we present an analog beamforming receiver architecture that uses multiple arrays of Phase Shifters and a single RF chain to combat the effect of angular errors, showing that it can achieve the same performance as hybrid beamforming.

 \end{abstract}

\section{Introduction and Related Work}
\label{sec:Rlt_wk}
In the past few years, there has been an increased interest in millimeter wave (mmWave) technology to fulfill the data rate requirements foreseen for the fifth generation cellular communication (5G) \cite{KhanF_mmWave}. 
However, frequencies in mmWave bands experience high path-loss, which in comparison to microwave bands may result in a significant coverage reduction when considering omnidirectional communication.
To overcome these coverage issues, beamforming at mmWave is an effective solution. 
Due to the small wavelengths at mmWave frequencies, a large number of antennas can be packed in a small space, and this allows to generate high gains and highly directional beams.

Beamforming solutions proposed for LTE are quite different from what is required for mmWave communication.
In LTE, initial access is performed using omnidirectional communication, and beamforming is used only when directional information is available after initial physical layer access.
However, in mmWave communication beamforming may be required even during the initial access/cell search process to overcome the coverage problems of omnidirectional communication with mmWaves.  

Recently, two different approaches have been considered for directional initial access/cell discovery.
Firstly, in \cite{CaponeFS15_CI_BF}, considering a HetNet scenario, context information (CI) regarding mobile station (MS) positioning is provided to the mmWave base station (BS) by the microwave BS. 
Based on this, the mmWave BS points its beam (using analog beamforming) in the desired direction. 
The authors also address the issue of erroneous CI, 
proposing that the BS, in addition to searching in the CI based direction, also searches the rest of the angular space by forming beams in different directions and also with different beamwidth (to increase the coverage).
Results showed that the enhanced cell discovery, where in case of positioning error the BS searches the adjacent angular directions, outperforms the greedy search approach where the BS searches the angular space sequentially. 
In addition, at the MS, omnidirectional reception is considered, which in comparison to a directional reception results in a reduced gain.
Recently, the authors of \cite{CaponeFS15_CI_BF} extended their work by considering a more complex channel model with multiple rays and obstacles \cite{Capone_ObsAv15}.    


Secondly, a non context information based approach is used in \cite{Barati_IBF, Alkhateeb_IBF, Marco_IniAcc16}.
In \cite{Barati_IBF}, omnidirectional and random directional transmissions are considered from the mmWave BS, and the MS can perform either analog, digital or hybrid beamforming.
It is observed that using an omnidirectional transmission at the BS and digital beamforming at the MS outperforms the other schemes.
However, the use of omnidirectional transmission at the BS during initial synchronization/access will reduce coverage, while random directional transmission will result in large delays.

In \cite{Alkhateeb_IBF}, exhaustive and hierarchical searches are compared while considering analog, digital and hybrid beamforming at the BS and the MS. 
In exhaustive search, the whole angular space is covered by sequentially transmitting beams in a time division multiplexing fashion, and initial beamforming is done by selecting the best combination of Tx-Rx beams. 
The hierarchical search, instead, is a multiple step process.
In the first step, a MS initially utilizes fewer antennas to form a relatively small number of wide beams.
The received signal is combined with all the beams and the best combiner beam is selected as a reference for the next step, where several narrower beamwidth combiners are formed, within the initially selected beam. 
Considering scenarios with limited mobility, the process finishes when the combiner beam is within the range of $5^\circ$ to $10^\circ$.
However, selecting an incorrect combiner in the initial stage can result in an initial access error in the following stage.

Recently, in \cite{Marco_IniAcc16}, iterative and exhaustive search schemes using analog beamforming have been studied and compared, and the authors showed that the optimal scheme depends on the target SNR regime.

\subsection{Our Approach}
\label{subsec:Our_app}
Analog beamforming (ABF) results in a lower power consumption compared to digital and hybrid beamforming (DBF and HBF) but has been shown to have comparable performance in terms of initial access probability when combined with an exhaustive search \cite{Alkhateeb_IBF}. 
Although the exhaustive approach is not preferred in general due to its inherent search delay,
the availability of CI can greatly improve its delay performance.  
To address this, in contrast to \cite{CaponeFS15_CI_BF}, we consider the availability of CI at the MS, so that the MS, rather than searching the whole angular space, will form its combining beam only in the direction provided by the CI\footnote{The initial cell search delay associated with the directional communication (for $N_{BS}=64$ and $N_{MS}=16$ with ABF) is roughly around 10 ms \cite{my_ICD16}, and is small enough to ingnore the rotational motion of the MS, which makes the directional signal reception at the MS a feasible option.}. 
This will result in a reduced initial cell search delay and, in comparison to omnidirectional reception (as in \cite{CaponeFS15_CI_BF}), the ABF at the MS will provide a higher gain.  
Moreover, the availability of CI at the BS does not necessarily minimize the delay, as if there are multiple MSs belonging to different beams the BS has to scan all of them, losing at least part of the delay savings. 
On the contrary, in initial cell search the MS typically listens to a single (or a small number of) BS, and therefore the availability of CI allows it to form a single beam in the right direction, thereby avoiding the beam scanning delay.

We consider a HetNet scenario 
for transferring the CI to the MS, where a microwave BS during the exchange of the initial control signals also transfers the location information of the mmWave BS (e.g., global positioning system (GPS) coordinates) to the MS. 
Due to the recent increase in location based applications and GPS positioning accuracy, GPS based CI may be considered as a viable option.
We assume that the GPS coordinates of the MS are already available. 
The MS, after acquiring the CI, figures out the expected angle of arrival (AoA) and aligns its beam in the required direction to receive the initial synchronization signals from the mmWave BS. 
In this work, our results give a quantitative evaluation of the benefit of having CI, by measuring how much better a mmWave system can perform with the availability of CI. 
A detailed assessment of the cost of obtaining the CI is beyond the scope of this paper, and is left as future work. 
 
 

 
Specifically, in this paper we address the following issues:
\begin{itemize}
\item how ABF with CI performs in comparison to ABF with non CI based approaches (Random \cite{Barati_IBF} and Exhaustive \cite{Alkhateeb_IBF} search);
\item how the angular error in the provided CI will affect the performance of the initial access process;
\item how the optimal number of MS antennas that results in the best access performance varies with the angular error in the available CI. 
\end{itemize}

Finally, we propose an analog beamforming based phase shifters network (PSN) architecture (see Figure \ref{fig:PSN}) to mitigate the effect of the angular error in the available CI, and also compare its performance with hybrid and digital beamforming in terms of initial access error and power consumption.

The rest of the paper is organized as follows. 
In Section \ref{sec:Sys_mod}, we define the MIMO based system with beamforming and extend the idea to CI and also present our PSN architecture. 
Next, we discuss the simulation results for the different mentioned beamforming schemes in Section \ref{sec:Sim_Res}, and finally conclude the paper in Section \ref{sec:Con}.   

\section{System Model}
 \label{sec:Sys_mod}
Consider a downlink mmWave MIMO communication system, with $N_{BS}$ BS antennas and $N_{MS}$ MS antennas. 
A typical MIMO received signal model, without considering beamforming, is given as \cite{rappaport2014millimeter} 
\begin{equation}
 \textbf{y} = \sqrt{P}\textbf{Hs}\ +\ \textbf{n} 
\end{equation}  
\normalsize 
where $\textbf{H}$ is $N_{MS} \times N_{BS}$ matrix which represents the channel between the BS and the MS, $\textbf{s}$ and $\textbf{y}$ are the transmitted and received symbol vectors, respectively, $P$ is the transmitted power, and $\bf n$ is the complex white Gaussian noise, $\textbf{n} \sim \mathcal{CN}(0,\sigma^2)$. 
The mmWave MIMO channel can be modeled with a few scatterers ($L$) \cite{KhanF_mmWave} and is well represented by the following geometric model \cite{Alkhateeb14_ChanEst} 
\begin{equation}
   \textbf{H} = \sqrt{\dfrac{N_{BS}N_{MS}}{\rho L}}\sum_{l=1}^{L}\eta_l\textbf{a}_{MS}(\phi_{l}) \textbf{a}_{BS}^H(\theta_{l})
\end{equation}
\normalsize
where $H$ represents the conjugate transpose, $\rho$ is the path-loss, $\eta_l$ is the complex gain associated with the $l^{th}$ path, $\textbf{a}_{MS}$ and $\textbf{a}_{BS}$ are the spatial signatures of the MS and the BS, respectively, and $\phi_l$ and $\theta_l$ $\in [0,2\pi)$ represent the AoA and angle of departure (AoD) of the $l^{th}$ path at the MS and the BS, respectively.
In this paper, for simplicity we restrict our analysis to single path line of sight (LOS) scenarios and therefore will not consider the subscript $l$ in further analysis, whereas the study of a multi-path scenario is left as future work.
Moreover, both the MS and the BS are equipped with a uniform linear array (ULA).
Now, considering the ULA at the BS, 
$\textbf{a}_{BS}$ is defined as
%
\begin{equation}
   \textbf{a}_{BS} = \dfrac{1}{\sqrt{N_{BS}}}[1, e^{j(2\pi /\lambda) d\sin (\theta)}, ... , e^{j(N_{BS} - 1)(2\pi /\lambda)d\sin (\theta)}]^T 
\end{equation}
where $T$ represents the transpose, $d$ is the spacing between antenna elements, and $\lambda$ is the wavelength of the transmitted signal. For $d = \lambda/2$, $\textbf{a}_{BS}$ becomes
\begin{equation}
   \textbf{a}_{BS} = \dfrac{1}{\sqrt{N_{BS}}}[1, e^{j{\pi}\sin (\theta)}, ... , e^{j(N_{BS} - 1){\pi}\sin (\theta)}]^T 
\end{equation}
\normalsize
and the spatial signature $\textbf{a}_{MS}$ can be defined similarly.

\subsection{Signal Model with Analog Beamforming}
\label{ABF}

The received signal, after applying beamforming and combining at the transmitter (the BS) and the receiver (the MS), can be written as  
\begin{equation}
   y = \sqrt{P}\textbf{w}_{MS}^H\textbf{H}\textbf{w}_{BS}{s}\ +\ \textbf{w}_{MS}^H\textbf{n}
   \label{eq:Tx-Rx_BF} 
\end{equation} 
where $\textbf{w}_{BS}$ and $\textbf{w}_{MS}$ are the transmit beamforming and receiver combining vectors, respectively. 
For simplicity in the notation, the subscripts for the BS and the MS are removed in our general description of beamforming vectors, however, we explicitly use them whenever necessary. 

\begin{figure}
        \centering
        \psfrag{Access Error Probability}{$p$}
        \scalebox{1}{\includegraphics[width=\columnwidth]{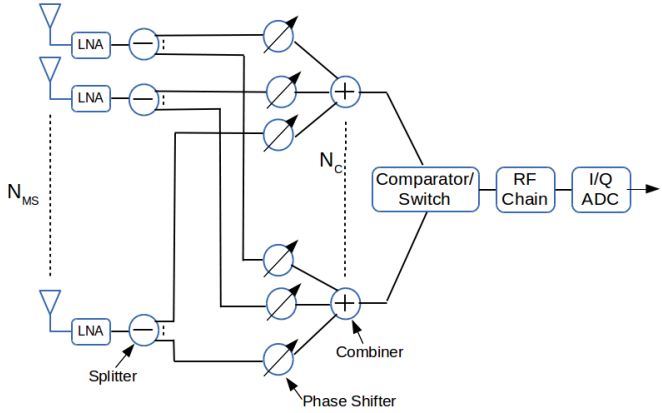}}\vspace{-4mm}
        \caption{\protect\renewcommand{\baselinestretch}{1}\footnotesize A Phase Shifters Network based receiver architecture.}\vspace{-4mm}
\label{fig:PSN}
 \end{figure}





Considering ABF, the beamforming vector $\textbf{w}$ models the analog phase shifters which apply successively progressive phases to the antenna elements. 
These phase shifters are digitally controlled and thus result in a finite number of possible phases.
Considering $q$ quantization bits, $2^q$ different phases can be applied to each element of an antenna array. 
The set of these quantized phases is represented as
\begin{equation}
  {t} = \bigg{\{}0,\ 2\pi(\frac{1}{2^q}),\ 2\pi(\frac{2}{2^q}),\ ...,\ 2\pi(\frac{2^q-1}{2^q})\bigg{\}} 
   \label{eq:Quan_Ph}
\end{equation} 
and with $N$ antennas, a codebook of $2^{qN}$ beamforming vectors can be generated.
However, in order to reduce the complexity, we only consider a subset of this codebook, and let $\textbf{W} = [\textbf{w}^1\ \textbf{w}^2\ \ldots\ \textbf{w}^{N_q}]$ represent an $N\times N_q$ reduced size codebook, where each column of $\textbf{W}$ represents a unique beamforming vector.   
Now for a quantized phase $\vartheta_i \in {t}$, we generate the receiver combining vector $\textbf{w}^i$ as
\begin{equation}
   \textbf{w}^i = \dfrac{1}{\sqrt{N}} [1,\ e^{j\vartheta_i},\ e^{j2\vartheta_i},\ ...,\ e^{j(N-1)\vartheta_i}]^T
   \label{eq:BF_vec}
\end{equation} 
where $i = 1,\ 2, \ldots\ N_q$.
Moreover, we set the cardinality of $\textbf{W}$ as $card(\textbf{W}) = N_q = 2N$, and therefore by definition of $\textbf{w}^i$ in (\ref{eq:BF_vec}), the required number of quantization bits is $q = \log _2 (2N)$.
This gives an acceptable performance, as shown in \cite{ABF_Codebook_Wang09}.
The transmitter and the receiver beamforming vectors 
are assumed to be defined according to the above approach, and $\textbf{W}_{MS}$ and $\textbf{W}_{BS}$ represent the beamforming codebooks at the MS and the BS, respectively.


\subsection{MS Combining Vector Selection with the Available CI}
We consider a scenario where the BS transmits sequentially in different angular directions using ABF, and the MS combines the received signal  with a combining vector selected based on the available CI. 
Note that the quantized phase $\vartheta_i$ is related to the physical AoA ($\phi$) as $\phi = \sin^{-1}(\vartheta_i/\pi)$.
The MS first estimates the CI based AoA ($\phi_{CI}$) and then identifies $\vartheta_i$ that minimizes the angular distance with $\phi_{CI}$. 
Finally, the combining vector that corresponds to the best $\vartheta_i$ is selected.
This combining vector maximizes the norm $|{\textbf{w}_{MS}^i}^H\textbf{a}_{MS}(\phi_{CI})|$, where $\textbf{a}_{MS}(\phi_{CI})$ represents the CI based estimated spatial signature at the MS. 
This maximization problem can be expressed as 

\begin{equation}
\begin{split}
   \textbf{w}_{MS}^\star =\ &\operatorname*{arg\,max}_{\textbf{w}_{MS}^i} |{\textbf{w}_{MS}^i}^H\textbf{a}_{MS}(\phi_{CI})| \\
   &s.t.\ {\textbf{w}_{MS}^i} \in \textbf{W}_{MS}
\end{split}   
   \label{eq:CI_dist}  
\end{equation}
the formulation of (\ref{eq:CI_dist}) also maximizes the signal to noise ratio ($SNR$) of the received signal (Equation (\ref{eq:Tx-Rx_BF})), expressed as
 
\begin{equation}
   SNR = \frac{|\textbf{w}_{MS}^H \textbf{H} \textbf{w}_{BS}|^2P}{||\textbf{w}_{MS}^H||^2\sigma^2}
   \label{eq:SNR}
\end{equation} 
\normalsize
where $P$ is the power of the transmitted signal, and $\sigma^2$ is the noise power.


We now extend this approach by considering an error in the available CI.
This will add an angular error $\phi_e$ to the estimated AoA, which may result in a wrong selection of $\vartheta_i$ and $\textbf{w}_{MS}$, which finally decreases the $SNR$ of the received signal.
We model $\phi_e$ as uniformly distributed in $[-\phi_e^{max}, \phi_e^{max}]$, where $\phi_e^{max}$ represents the maximum angular error\footnote{The general trends followed by different beamforming schemes (as shown in Section \ref{sec:Sim_Res}) will be maintained if different angular error distributions are used.}.
The amount by which $\phi_e$ affects the selection of $\vartheta_i$ is related to the 3 dB beamwidth $\theta_{BW}$ of the antenna array,
which for a ULA with $ d = \lambda/2$ and $N$ antennas can be computed as $\theta_{BW} = 2\sin^{-1}(0.891/N)$ \cite{oap-b1105852}. 
Therefore, for a reliable performance it is desirable to have $\phi_e^{max} \leq \theta_{BW}$. 
However, to address larger $\phi_e^{max}$, i.e., $\phi_e^{max} > \theta_{BW}$, one possible solution is to increase $\theta_{BW}$ by reducing the number of antennas at the cost of a reduced gain. The other option is to form multiple beams simultaneously, as discussed in the next section.

\subsection{Phase Shifters Network to Combat $\phi_e$ }
\label{ssec:PSN}
The performance of ABF (which only results in a single beam) starts degrading as $\phi_e^{max}$ becomes comparable to $\theta_{BW}$.
A solution to mitigate the effect of large $\phi_e^{max}$ is to form multiple simultaneous beams and then identify among them the best combining vector.
HBF or DBF  are attractive options to form multiple simultaneous beams\footnote{For details regarding HBF and DBF, see \cite{Alkhateeb14_ChanEst} and \cite{Litva:1996:DBW:547927}, respectively.}, but at the cost of higher power consumption. 
Compared to a fully Digital architecture (DBF), HBF is preferred for mmWave communication due to its lower power consumption because of its reduced number of RF chains. 
However, the required number of ADCs in HBF, which are considered as the main power hungry blocks in mmWave receiver design, increases proportionally to the number of RF chains.
Therefore, in comparison to ABF, HBF and DBF may result in significantly higher power consumption\footnote{However, a more detailed power consumption analysis that takes into account all components in addition to the ADCs reveals that there are interesting tradeoffs, and there exist regimes in which DBF may actually be a more convenient choice than HBF and in some cases even almost as good as ABF \cite{My_PCComp_EW16}.}.

To jointly address the issue of higher power consumption and large $\phi_e^{max}$, we propose a phase shifters network (PSN) architecture.
The architecture of PSN is similar to ABF but instead of forming a single beam with one combining vector, PSN allows the formation of multiple beams simultaneously by using multiple combining vectors.
The idea is to form multiple beams and identify the best combining vector which corresponds to the desired AoA all in the analog domain.
Therefore, PSN provides a power efficient design with only two ADCs (for the inphase and the quadrature phase signals), whereas in HBF the number of ADCs is directly proportional to the number of RF chains. 
The PSN receiver architecture is shown in Figure \ref{fig:PSN}, where $N_C$ is the number of combiners.
The number of simultaneous beams is equal to $N_C$, where phase shifters connected to a particular combiner represent a unique receiver combining vector.
The received signal is combined with $N_C$ combining vectors and the output of each combiner is compared using a comparator and with the help of a switch only the output of the best combiner (the one with the strongest signal) is forwarded to the ADC for further digital processing.

The architecture of PSN is also similar to HBF, with the exception that it consists of a single RF chain and allows beamforming only in the analog domain.
Therefore, in comparison to HBF and DBF,  PSN does not provide the advantages of digital beamforming.
However, during cell search the inherent advantages of digital beamforming like multiplexing, interference cancellation or multiuser communication are not required at the MS\footnote{ Note that, in order to get the highest energy efficiency in the first phase of initial access while enjoying the advantages of digital beamforming in subsequent phases and during data communication, the PSN architecture can be easily converted to HBF by switching the output of each combiner to a separate RF chain rather than to the comparator. A detailed study of such hybrid architecture is part of our future work.}.     
Therefore, PSN, thanks to its lower power consumption receiver design, is a viable option for mmWave initial cell search.

To form multiple beams, the selection of the main combining vector for PSN follows a similar formulation as in Equation (\ref{eq:CI_dist}).
However, in contrast to ABF where only a single combining vector is required, to form multiple beams in PSN $N_C$ different combining vectors from codebook $\textbf{W}_{MS}$ must be used.
Among the $N_C$ different combining vector, the main combining vector is selected according to (\ref{eq:CI_dist}), while the other $N_C-1$ combining vectors are selected from $\textbf{W}_{MS}$ such that they have minimum angular distance from the main combining vector.
According to the formation of the combining matrix $\textbf{W}_{MS}$, this results in a selection of $N_C-1$ combining vectors that are adjacent to the main combining vector.
This is a suitable choice as $\textbf{W}_{MS}$ is generated in a way to ensure that the gain fluctuation among any two adjacent combining vectors is within 1 dB \cite{ABF_Codebook_Wang09}. 

\section{Simulation Setup and Results}
\label{sec:Sim_Res}

In our performance evaluation, we assume that the CI of the mmWave BS is 
already available at the MS and with this location information the MS can point its beam in the desired direction.  
We also assume that the mmWave BS always transmits the signal in different directions sequentially using ABF, 
while for a better comparison different beamforming schemes are considered at the MS. 
A more comprehensive comparison considering different beamforming options also at the BS is left for future study.
%
  
In our simulations, we consider a single path LOS scenario $(L = 1)$, Rician fading with $k = 10$, carrier frequency of $28$ GHz, transmit power of $30$ dBm, noise figure of $5$ dB, thermal noise power of $-174$ dBm/Hz, path-loss exponent of $2.2$, 
and 64 transmit antennas at the BS.
The simulation results are averaged over $10^5$ different channel realizations for each Tx-Rx distance. 
We evaluate the performance in terms of access error probability $P_{AccEr}$, which is defined as the probability that the $SNR$ of the received signal at the MS (Eq. (\ref{eq:SNR})) is below a certain threshold $\Theta$.
\begin{equation}
   P_{AccEr} =  \Pr(SNR < \Theta)
\end{equation}
In our simulations this threshold $\Theta$ is set to $-4$ dB \cite{Alkhateeb_IBF}.

\begin{figure}
        \centering
        \psfrag{Access Error Probability }{\ \ \ \ \ \ \ \ $P_{AccEr}$}
        \psfrag{Tx-Rx Distance }{\back\ $Tx-Rx Distance$}
        \psfrag{Nbs = 64}{\back\back\back $N_{BS} = 64, \phi_e = 0^\circ$}
        \scalebox{1}{\includegraphics[width=\columnwidth]{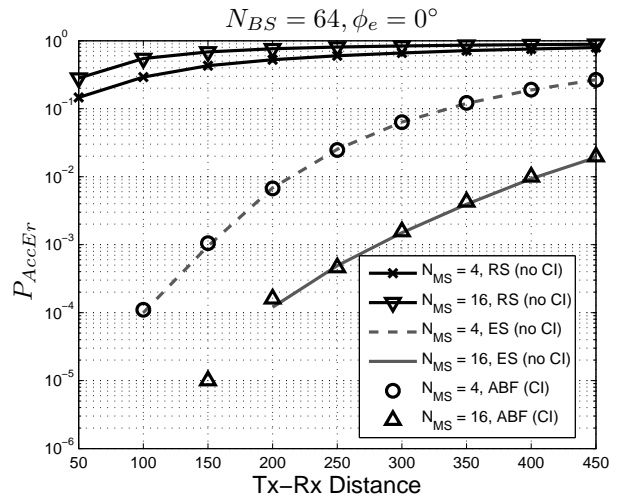}}\vspace{-4mm}
        \caption{\protect\renewcommand{\baselinestretch}{1.25}\footnotesize A comparison of Random Search, Exhaustive Search, and CI based schemes. }\vspace{-4mm}
\label{fig:ABF_CI_vs_noCI}
\end{figure}

We start with a comparison of our proposed CI (with ABF) based initial cell search scheme with two non CI (with ABF) approaches, namely Random search (RS) \cite{Barati_IBF} and Exhaustive search (ES) \cite{Alkhateeb_IBF}. 
Figure \ref{fig:ABF_CI_vs_noCI} shows how $P_{AccEr}$ for CI and non CI based approaches varies for different values of Tx-Rx Distance and $N_{MS}$. 
Results show that RS performs worst in comparison to the other two schemes. 
This is because in RS a MS forms its beam in a randomly selected direction, and therefore it is less likely to form a beam in the required direction, which results in a higher initial access error.
Also in RS, as $N_{MS}$ increases, the beamwidth gets narrower and the MS has to form more beams to cover the complete angular space, and therefore the probability of finding the right combining beam decreases, which results in an increase in $P_{AccEr}$.  
On the other hand ES, where a MS searches the whole angular space sequentially, and our proposed CI based initial cell search approach show similar performance. 
However, considering ABF both at the BS and the MS, the exhaustive approach has a large search delay because $card(\textbf{W}_{BS}) \times card(\textbf{W}_{MS})$ joint angular directions need to be considered, whereas for our CI based approach this search delay is reduced to $card(\textbf{W}_{BS}) \times 1$, at the cost of obtaining CI.
For the rest of this section, the performance evaluation is only focused on the CI based approach.
We now discuss how the CI based approach performs in the presence of the angular error in the available CI, and also how this angular error is related to $N_{MS}$.

\begin{figure}[t]
     \psfrag{Access Error Probability }{\ \ \ \ \ \ \ \ \ $P_{AccEr}$}
     \psfrag{AnEr = 0A}{\scriptsize $\phi_e^{max} = 0^{\circ}$, $N_{MS} = 4$}
          \psfrag{AnEr = 5A}{\scriptsize $\phi_e^{max} = 5^{\circ}$, $N_{MS} = 4$}
               \psfrag{AnEr = 10A}{\scriptsize $\phi_e^{max} = 10^{\circ}$, $N_{MS} = 4$}
                \psfrag{AnEr = 0P}{\scriptsize $\phi_e^{max} = 0^{\circ}$, $N_{MS} = 16$}
          \psfrag{AnEr = 5P}{\scriptsize $\phi_e^{max} = 5^{\circ} $, $N_{MS} = 16$}
               \psfrag{AnEr = 10P p p p p p p p p p p p p}{\scriptsize $\phi_e^{max} = 10^{\circ}$, $N_{MS} = 16$}
               \psfrag{AnEr = 10 PSN Nc3}{\scriptsize $\phi_e^{max} = 10^{\circ}, N_{MS} = 16 $}
        \scalebox{1}{\includegraphics[width=\columnwidth]{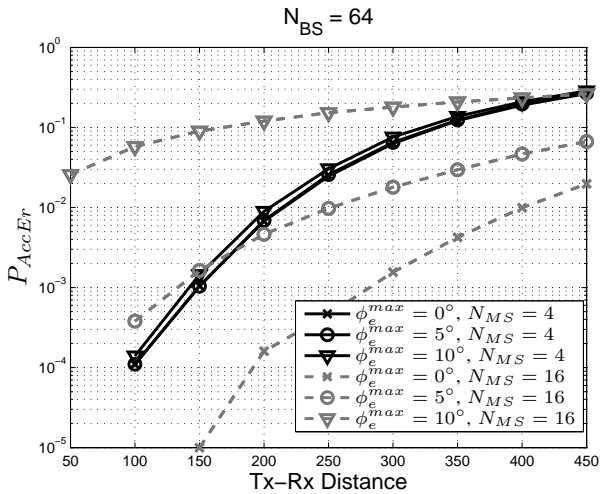}}\vspace{-4mm}
        \caption{\protect\renewcommand{\baselinestretch}{1.25}\footnotesize Access error probability vs Tx-Rx distance for ABF with CI for different values of $N_{MS}$ and with $\phi_e^{max}$.}\vspace{-4mm}
        \label{fig:ABFNms4vs16}%
\end{figure}

In Figure \ref{fig:ABFNms4vs16}, $P_{AccEr}$ is plotted for different values of $\phi_e^{max}$, while considering 4 and 16 antennas at the MS, respectively. 
As expected, for $\phi_e^{max} = 0$, with an increase in $N_{MS}$ the beamforming gain increases, and therefore $P_{AccEr}$ decreases.  
For both $N_{MS} = 4$ and $N_{MS} = 16$, $P_{AccEr}$ increases with an increase in the Tx-Rx distance or the angular error $\phi_e$.
The increase in $P_{AccEr}$ with an increase in $\phi_e^{max}$ for $N_{MS} = 4$ is not very significant due to its wider beamwidth ($\theta^4_{BW} = 25.7^\circ $). 
However, in the case of $N_{MS} = 16$, 
the beamwidth decreases ($\theta^{16}_{BW} = 6.38^\circ $), and  
therefore with an increase in the angular error it is more likely that the incoming signal will not fall within $\theta_{BW}$, and this results in a significant increase in $P_{AccEr}$.

Figure \ref{fig:ABFNms4vs16} also highlights how the angular error is related to the optimal number of receive antennas.
It is observed that when the angular error starts falling outside of $\theta_{BW}$ of the estimated beam, $P_{AccEr}$ also increases.
Therefore, an increase in $N_{MS}$ makes it more susceptible to angular errors and hence results in even worse performance.
It is obvious from Figure \ref{fig:ABFNms4vs16} that the optimal number of antennas varies with $\phi_e^{max}$. 
For instance, with $\phi_e^{max} = 0^\circ$, $N_{MS} = 16$ is the better option, whereas for $\phi_e^{max} = 10^\circ$, $N_{MS} = 4$ performs better than $N_{MS} = 16$.
Moreover, for $\phi_e^{max} = 5^\circ$, $N_{MS} = 16$ has a lower $P_{AccEr}$ for Tx-Rx distance 50 m and above, whereas below 50 m $N_{MS} = 4$ is a preferable choice.
This shows a tradeoff between distance dependent path-loss and $\phi_e$. 
At short Tx-Rx distance, $\phi_e$ has more significance on $P_{AccEr}$ and therefore $N_{MS} = 4$ performs better, however as the Tx-Rx distance increases the path-loss also increases and therefore $N_{MS} = 16$ with higher beamforming gain results in a lower $P_{AccEr}$.
  


\begin{figure}
        \centering
        \psfrag{Access Error Probability }{\ \ \ \ \ \ \ \ \ $P_{AccEr}$}
        \psfrag{Angular Error = 0 ABF}{\scriptsize$\phi_e^{max} = 0^\circ$ ABF}
        \psfrag{Angular Error = 5 ABF}{\scriptsize$\phi_e^{max} = 5^\circ$ ABF}
        \psfrag{Angular Error = 10 ABF}{\scriptsize$\phi_e^{max} = 10^\circ$ ABF}
        \psfrag{Angular Error = 0 PSN Nc = 3}{\scriptsize$\phi_e^{max} = 0^\circ$ PSN}
        \psfrag{Angular Error = 5 PSN Nc = 3}{\scriptsize$\phi_e^{max} = 5^\circ$ PSN}
        \psfrag{Angular Error = 10 PSN Nc = 3}{\scriptsize$\phi_e^{max} = 10^\circ$ PSN}
       
        \scalebox{1}{\includegraphics[width=\columnwidth]{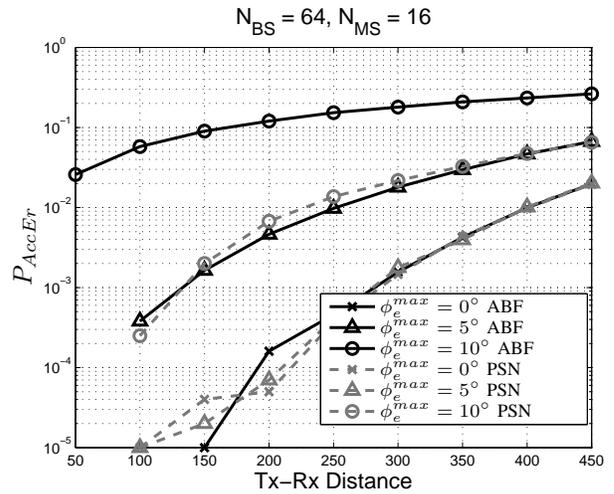}}\vspace{-4mm}
        \caption{\protect\renewcommand{\baselinestretch}{1.25}\footnotesize Access error probability vs Tx-Rx distance for ABF and PSN with $N_C = 3$. }\vspace{-1mm}
\label{fig:ABFvsPSN_Nms16}
\end{figure}

We now compare the performance of PSN with ABF and show how PSN can mitigate the effect of $\phi_e$.
For PSN, we consider $N_C = 3$. 
Figure \ref{fig:ABFvsPSN_Nms16} presents a comparison of ABF and PSN for different angular errors. 
For both schemes, $P_{AccEr}$ increases with an increase in $\phi_e^{max}$ or Tx-Rx distance.    
However, PSN performs better than ABF, and is less prone to a small increase in $\phi_e^{max}$.
For instance, the performance of PSN is unaffected for $\phi_e^{max} = 5^\circ$, whereas there is an increase in $P_{AccEr}$ for $\phi_e^{max} = 10^\circ$. 
This is because the PSN architecture allows to form multiple simultaneous beams and is able to capture the signal energy even in the presence of angular errors.
The performance of PSN in case of angular error improves with $N_C$, as a 
larger $N_C$ will result in more combiner beams covering more angular space and thus is less susceptible to angular errors.

As PSN outperforms ABF in case of erroneous CI, we now further evaluate its performance by comparing it with HBF and DBF. In simulations, for PSN and HBF $N_C = N_{RF} = 3$. 


\begin{figure}
        \centering
        \psfrag{Access Error Probability}{\ \ \ \ \ \ \ \ \ $P_{AccEr}$}
         \psfrag{Angular Err = +/-0, PSN}{\scriptsize$\phi_e^{max} = 0^\circ$, PSN}
       \psfrag{Angular Err = +/-10, PSN}{\scriptsize$\phi_e^{max} = 10^\circ$, PSN}
        \psfrag{Angular Err = +/-0, DBF}{\scriptsize$\phi_e^{max} = 0^\circ$, DBF}
        \psfrag{Angular Err = +/-10, DBF}{\scriptsize$\phi_e^{max} = 10^\circ$, DBF}
        \psfrag{Angular Err = +/-0, HBF}{\scriptsize$\phi_e^{max} = 0^\circ$, HBF}
        \psfrag{Angular Err = +/-10, HBF}{\scriptsize$\phi_e^{max} = 10^\circ$, HBF}
        \scalebox{1}{\includegraphics[width=\columnwidth]{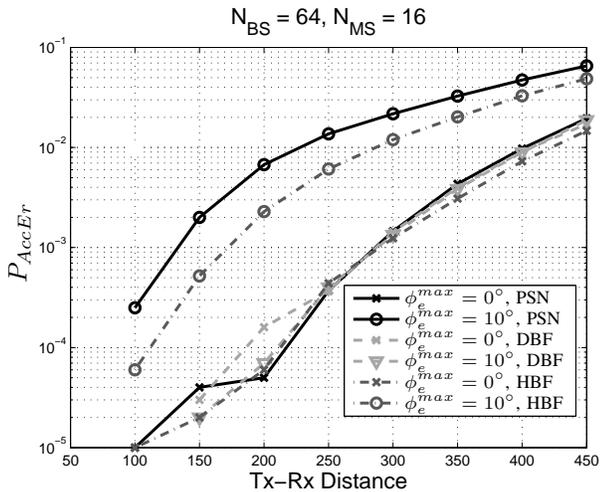}}\vspace{-4mm}
        \caption{\protect\renewcommand{\baselinestretch}{1.25}\footnotesize Access error probability vs Tx-Rx distance for PSN, HBF and DBF.}\vspace{-5mm}
\label{fig:DBFvsHBFvsPSN}
\end{figure}

Figure \ref{fig:DBFvsHBFvsPSN} presents a comparison of DBF and HBF with PSN.
It is depicted that DBF shows a stable performance with an increase in the angular error. 
This is because DBF allows to form multiple beams which allows the MS to look in all angular directions simultaneously, and therefore the performance of DBF is not effected by angular errors (but at the cost of higher power consumption). 
Note that the performance of PSN and HBF for small angular errors is similar to that of DBF.

Moreover, for $\phi_e^{max} = 0^{\circ}$, HBF results in a slightly lower $P_{AccEr}$ than PSN due its additional digital baseband combiners (which further minimizes the distance between $\phi_{CI}$ and $\vartheta_{i}$), but the performance difference between PSN and HBF diminishes with an increase in $\phi_e^{max}$. 
Also, note that the digital baseband combiner in HBF requires additional RF chains, which results in higher power consumption. 
Therefore, PSN is a viable solution as it provides significant power savings at the cost of only a small performance degradation at lower $\phi_e^{max}$.

In order to make the above considerations about power consumption of the various beamforming schemes more precise, we study how the total receiver power consumption $P_{Tot}$, varies with an increase in the number of ADC bits for ABF, DBF, HBF and PSN. 
In our evaluation, we consider $N_C = N_{RF} = 3$ and $N_{MS} = 16$. 
$P_{Tot}$ for ABF and DBF can be calculated similar to \cite{OrhanER15_PowerCons}\footnote{For a detailed power consumption comparison among ABF, DBF and HBF, the reader is referred to \cite{My_PCComp_EW16}.}   
\begin{equation}
  P_{Tot}^{ABF} =  N_{MS}(P_{LNA} + P_{PS}) + P_C + P_{RF} + 2P_{ADC}
   \label{eq:ABF}
\end{equation}    
\begin{equation}
  P_{Tot}^{DBF} =  N_{MS}(P_{LNA} + P_{RF} + 2P_{ADC})
   \label{eq:DBF}
\end{equation}  
and similarly $P_{Tot}$ for HBF \cite{Alkhateeb15_SwvsPS} and PSN can be calculated as
\begin{equation}
\begin{split}
    P_{Tot}^{HBF} = &\ N_{MS}(P_{LNA} + P_{SP} + N_{RF}P_{PS}) \\			& + N_{RF}(P_C + P_{RF} + 2P_{ADC})
\end{split}
\label{eq:HBF}   
\end{equation}   
\begin{equation}
\begin{split}
    P_{Tot}^{PSN} &=\ N_{MS}(P_{LNA} + P_{SP} + N_{C}P_{PS}) \\
   & + N_{C}P_C + P_{RF} + P_{Comp} + P_{Sw} + 2P_{ADC}
\end{split}
\label{eq:PSN}
\end{equation} 
where $P_{RF} = P_{M} + P_{LO} + P_{LPF} + P_{BB_{amp}}$ represents the power consumption of the RF chain, and 
where $P_{LNA}$, $P_{PS}$, $P_C$, $P_M$, $P_{LO}$, $P_{LPF}$, $P_{BB_{amp}}$, $P_{ADC}$, $P_ {SP}$, $P_{Sw}$ and $P_{Comp}$ represent the power consumption of low noise amplifier (LNA), phase shifter, combiner, mixer, local oscillator, low pass filter, baseband amplifier, ADC, splitter, switch and comparator, respectively. 
The power consumption of an ADC scales exponentially with the number of bits and linearly with the sampling rate \cite{ADC_b_B}. Therefore, considering Nyquist sampling rate, the ADC power consumption is modeled as
\begin{equation}
    P_{ADC} = cB2^b
\label{eq:Padc}
\end{equation}
where $b$ is the number of ADC bits, $B$ is the sampling rate (Bandwidth) and $c$ is the energy consumption per conversion step. 
For $B = 500$ MHz and $b = 5$, the power consumption of an ADC in \cite{Alkhateeb15_SwvsPS} is considered as 200 mW, and this correponds to $c = 12.5$ pJ. 
For details regarding the power consumption of other components, the reader is referred to \cite{Alkhateeb15_SwvsPS}.     

\begin{figure}
        \centering
        \psfrag{Total Power Consumption (Watts)}{\ \ \ \ \ \ \ \ \ \ $P_{Tot}\ (Watts)$}
        \psfrag{Number of ADC bits}{\ \ \ \ \ \ \ \ \ \ $b$}
        \scalebox{1}{\includegraphics[width=\columnwidth]{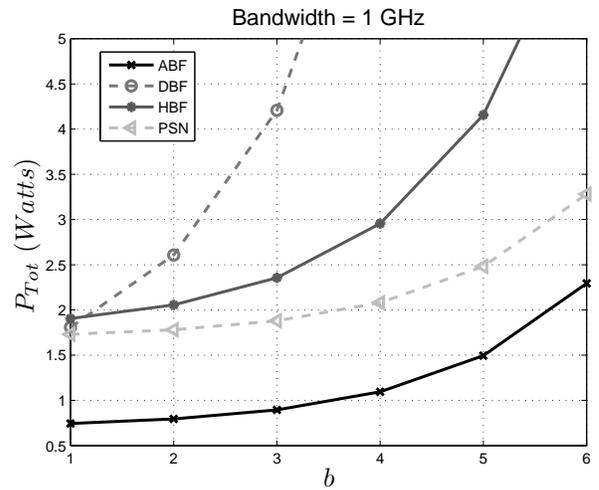}}\vspace{-4mm}
        \caption{\protect\renewcommand{\baselinestretch}{1.25}\footnotesize Power consumption comparison of ABF, DBF, HBF and PSN.}
\label{fig:PowerCons_Nms16}
\end{figure}
  
Figure \ref{fig:PowerCons_Nms16} shows that with an increase of $b$ there is an exponential increase in power consumption. 
As expected, $P_{Tot}$ for ABF is always lower than for the other beamforming schemes, and hence ABF with CI is a suitable option for initial access considering small angular errors.
On the other hand, DBF always has higher power consumption than other beamforming schemes except for $b = 1$.      
Finally, PSN consumes lower power than HBF irrespective of the number of ADC bits.
Moreover, the $P_{Tot}$ difference for PSN and HBF increases with an increase in $b$. 
Therefore, PSN with $P_{Tot}$ less than DBF and HBF is a viable option even for high resolution (i.e., higher number of bits) ADC based mmWave receiver design.


\section{Conclusions}
\label{sec:Con}

In this work, we discussed how the availability of CI at the MS can reduce the inherent search delay of ABF. 
However, the performance of ABF starts degrading with an increase of the angular error in the available CI.
We also showed that the optimal number of receiver antennas is related to the angular error. 
In addition, considering small angular errors in CI, ABF with lower power consumption is the best option for initial access.   
Moreover, we presented an analog beamforming based PSN architecture with a single RF chain to mitigate the effect of the angular error.
Simulation results validate that this solution has equivalent performance to HBF, while exhibiting lower power consumption.
This makes PSN a viable approach for initial cell search in mmWave 5G cellular networks.    

In the future, we will extend this work to multipath scenarios, and study how different beamforming schemes perform in the presence of multiple transmitting BS.
We will also evaluate the optimal number of receiver antennas based on the statistics of the angular error. 
Moreover, we will analyze the performance of PSN with respect to HBF and DBF without the availability of context information.  
Finally, we will study the choice of an appropriate beamforming scheme which jointly minimizes the access error probability and the energy consumption.

\bibliographystyle{IEEEtran}
\bibliography{IEEEabrv,biblio,refen}

\end{document}